\documentclass[referee]{aa501} 
\usepackage{graphicx}

\begin{document}
    \title{Cosmological constant and the fate of the DDM theory}

 \offprints{M. M. \'Cirkovi\'c}

\author{\bf Milan M. \'Cirkovi\'c $^{(1)}$ and Srdjan Samurovi\'c $^{(2)}$} 

\institute{$^{(1)}$   Astronomical Observatory, Volgina 7, 11000
Belgrade, YUGOSLAVIA \\
{\tt mcirkovic@aob.aob.bg.ac.yu}\\
$^{(2)}${Dipartimento di Astronomia, 
Universit\`a di Trieste, 
Via Tiepolo 11,
I-34131 Trieste, ITALY\\
{\tt srdjan@ts.astro.it}}
}

\date{Received  ; accepted  }

\abstract{
We investigate the impact of the non-zero cosmological constant on
the classical decaying dark matter theory developed by the late Dennis
Sciama. In particular, we concentrate on the change in relevant values of
cosmological parameters in comparison to the high-precision
estimates given by Sciama (1997). It is shown that the appropriate changes
in resulting parameter values are such to make the DDM concept less
plausible. This is in complete agreement with recently reported
observational results detrimental to this theory.
\keywords{elementary particles ---  cosmology: dark matter 
---  cosmology: diffuse radiation -- distance scale
               }
}

\titlerunning{Cosmological constant and  the DDM theory }
\authorrunning{\'Cirkovi\'c  \& Samurovi\'c}

\maketitle
\section{Introduction}

Decaying dark matter (DDM) theory is an attempt to simultaneously solve two
important problems of contemporary astrophysics: the dark matter problem in spiral
galaxies, like the Milky Way, and  the problem of ionization of the
interstellar and intergalactic
medium (Sciama 1993). To achieve these goals,  theory introduces massive decaying
neutrino with the mass $m_\nu \sim$ 30 eV. This neutrino has a decay lifetime of
$2\pm 1\times 10^{23}$ s, that produces a decay photon of energy of $13.7\pm 0.1$ eV
(Sciama 1998). This theory is heavily constrained, i.e. its parameters are very well
defined, with extremely small uncertainties.  An experiment, EURD,
has been proposed in order to test the theory (Sciama 1993). Results recently
published  suggest that this theory is no longer viable, because the emission predicted
by the DDM theory was not registered
(Bowyer et al. 1999). In this {\it Letter\/} we wish to investigate the values of the Hubble
constant and predicted age of the universe in the DDM theory, in the light of two
recent important empirical discoveries: first that
neutrinos do have mass (Fu\-ku\-da et al.\ 1998), and the second one according to which there
exists a large positive cosmological constant (Perlmutter et al.~1998, 1999;
Reiss et al.~1998).

In this respect, it seems that we are in the middle
of a major change of cosmological paradigm (not unexpected, however, as even the cursory
look at the relevant literature could show). Recent results of the surveys of the Type I
supernovae at cosmological distances indicate the possible presence of a large cosmological constant
(Perlmutter et al.~1998, 1999; Reiss et al.~1998). If the total
cosmological density parameter corresponds to the flat ($\Omega=1$) universe, the
contribution due to matter density is (total 1$\sigma$ statistical +
systematic errors quoted)
\begin{equation}
\label{jedan}
\Omega_m = 0.28^{+0.14}_{-0.12}.
\end{equation}
This result suggests not only that the universe will expand indefinitely, but that it will
expand in an (asymptotically) exponential manner,
manner, similar to the early inflationary phase in its history.
In addition to these observations, we use results from the
primordial nucleosynthesis which are entering the high-precision
phase (Schramm and Turner 1998), and limit the combination of baryonic density
fraction and the Hubble parameter. We shall use the following
(conservative) limits:
\begin{equation}
\label{dva}
\Omega_b h^2= 0.025 \pm 0.005.
\end{equation}
These are larger values than those used by Sciama (1997), but this
can be justified on several counts. First of all, later
measurements of deuterium abundance at high redshift unambiguously
indicate lower abundances than previous controversial values
(Burles and Tytler 1998). In addition, measurements of HeII
Gunn-Peterson effect at high redshift (Jakobsen 1998) gave very high values for $\Omega_b
h^2$, even higher than those in Eq.~(\ref{dva}). For the sake of
completeness, we have used both this realistic, and the lower value
of Sciama (1997) in further calculations.

One should add the following epistemological consideration.
Being the property of the quantum vacuum itself, addition of the non-zero cosmological
constant does not {\it prima facie\/} increase the conceptual
complexity of the theory for dark matter. However, if we believe
in classical prediction of the inflationary scenario $\Omega = 1 \pm \epsilon$
with the precision $\epsilon \simeq 10^{-5}$, we have to take into
account this additional constraint on the distribution of total energy
density in the universe. We shall use this assumption in the
further considerations.

We shall use the following notation: symbol $\Omega$ without any subscripts will be reserved
for the total density parameter of the universe, which, according to our present
understanding can be written as the sum of densities of
matter\footnote{Including presently negligible contribution of radiation
$\Omega_{\rm rad} =4.31 \times 10^{-5}\, h^{-2}$.}
and vacuum density (which is manifested in the form of the cosmological constant
$\Lambda$), i.e.
\begin{equation}
\label{tot}
\Omega \equiv \Omega_m + \Omega_\Lambda.
\end{equation}
The contribution of matter can be written as
\begin{equation}
\label{pirvo}
\Omega_m \equiv \frac{\rho_m}{\rho_{\rm crit}} = \frac{8 \pi G \rho_m}{3 H_0^2},
\end{equation}
and the one of the cosmological constant as
\begin{equation}
\label{bazar}
\Omega_\Lambda = \frac{c^2 \Lambda}{3H_0^2} = 2.8513 \times 10^{55} \, h^{-2} \Lambda,
\end{equation}
$\Lambda$ being in units of cm$^{-2}$. The present day Hubble
constant is pa\-ra\-met\-riz\-ed in a standard way as
$H_0 \equiv 100 \, h \;  {\rm km \; s}^{-1} \; {\rm Mpc}^{-1}$.
$\Lambda$ enters the Einstein field equations as
\begin{equation}
\label{polik}
R_{\mu \nu} - \frac{1}{2} g_{\mu \nu} R - \Lambda g_{\mu \nu} =
- \frac{8\pi G}{c^4} T_{\mu \nu},
\end{equation}
and $\Lambda$-universes are the homogeneous and isotropic
solutions of these tensor equations (for other notation see any of
the standard General Relativity textbooks, e.g.\ Weinberg 1972;
for history and phenomenology of the cosmological constant,
see the detailed review of Carroll, Press \& Turner 1992, and
references therein). We now wish to investigate whether an inflationary
DDM universe can be reconciled with non-zero $\Lambda$ and still
perform its explanatory tasks.

\section{Simple parameter estimates}

It is not possible to proceed in as simple way as in Sciama (1997),
since the combination $\Omega_\Lambda h^2$ does not have an obvious
physical meaning. However, we shall use this circumstance in order
to establish plausible values for $h$ first. For establishing connection between
ionizing flux $F$ and $m_\nu$ we follow the same
simple procedure
outlined by Sciama (1997), except that it is not possible any more to simply plug
in the "final" value for
$h$ as it has been done in that study. This situation gives
rise to the term linear in $h$, which is the main source of
difficulties here. Therefore, we obtain for the decaying neutrino
mass
\begin{equation}
\label{neut0}
m_\nu = (27.2 + 0.39 h) \pm 0.39 h \; {\rm eV}.
\end{equation}
Hence, the contribution to the cosmological density fraction
is
\begin{equation}
\label{neut1}
\Omega_\nu h^2= \frac{27.2 + 0.39 h}{93.6} = 0.2906 + 0.0042
h.
\end{equation}
Assuming in the spirit of DDM theory that $\Omega_m = \Omega_\nu +
\Omega_b$, we can write
\begin{equation}
\label{neut2}
\Omega h^2 = \Omega_\Lambda h^2+ \Omega_\nu h^2 + \Omega_b h^2,
\end{equation}
or, equivalently, taking into account Eqs.~(\ref{dva}) and (\ref{neut1}),
we have
\begin{equation}
\label{neut3}
(\Omega - \Omega_\Lambda) h^2 = 0.025 + 0.2906 + 0.0042 \,
h = 0.3156 + 0.0042\, h.
\end{equation}
Now we may use  the theoretical prejudice for $\Omega =1$
that is in agreement with the recent Boomerang result (de Bernardis et al. 2000), and
therefore $\Omega_\Lambda$ is determined by Eq.~(\ref{jedan}).
Later we shall discuss the consequences of variations in $\Omega$
in the observationally allowed range (approximately 0.3$-$1.1).
The physically acceptable solution of this quadratic equation in $h$ is
$h = 1.07$.

\begin{figure*}

{\includegraphics[height=9cm,angle=270]{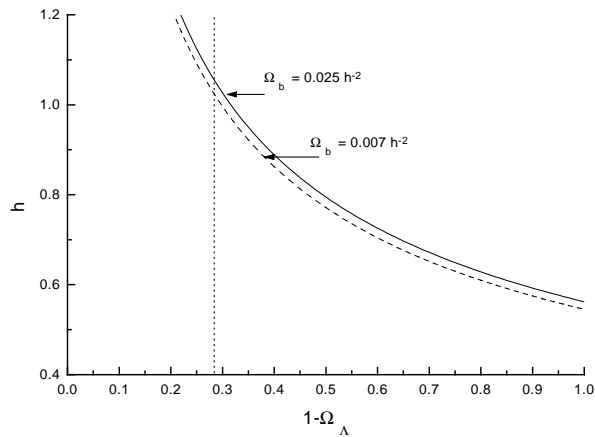}}
\vspace{0.5cm}

\caption{The value of the Hubble parameter, $h$, as a function
of the cosmological constant $\Lambda$,
i.e. its contribution to the total cosmological density (assumed
to be unity).  For the sake of clarity, we plotted $1-\Omega _\Lambda$
which represents contribution of the neutrinos and baryonic matter
to the total cosmological density. Vertical dotted line, plotted
at  $1 - \Omega_\Lambda \approx 0.28$ corresponds to the realistic
contribution of matter
in the universe. It is obvious that in this case $h$ tends to
the unrealistically high value of $h\sim 1$. The different possible contributions of
the baryonic matter to the total cosmological density are represented
by two curves: solid (realistic higher $\Omega_b$, see text) and dashed
(used by Sciama [1997]). \label{auto1}}

\end{figure*}

The physical picture here is highly intuitive: for the fixed total
density parameter, introduction of a term with negative effective
pressure in the Friedmann equation results in a faster expansion
rate. However, all recent observational measurements have suggested
lower values for the Hubble parameter, in the 0.5 -- 0.8 range,
even tending toward the lower limit (e.g. Paturel et al.~1998; Schaefer
1998). We notice that we recover the particular value $h = 0.55$
for $\Lambda=0$ obtained by Sciama (1997), as expected.

\section{Discussion}

The age of the universe predicted in such any theory with non-zero $\Lambda$ is given as
(e.g. Weinberg 1972; Carrol, Press and Turner 1992)
\begin{equation}
\label{cpt}
t_0= \frac{1}{H_0} \int_0^1dx [(1-\Omega_m - \Omega_\Lambda) +
\Omega_m x^{2-3(1+ \alpha)} + \Omega_\Lambda x^2]^{- \frac{1}{2}}.
\end{equation}
Here $\alpha$ defines the equation of state of the present matter,
being effectively the ratio of pressure to energy density. For $\Omega =1$
case, this reduces to the well-known relation (e.g. Singh 1995)
\begin{equation}
\label{cpt2}
t_0 = \frac{2}{3 H_0} \frac{1}{\sqrt{\Omega_\Lambda}} \ln
\left[\frac{1+ \sqrt{\Omega_\Lambda}}{\sqrt{1- \Omega_\Lambda}}.
\right]
\end{equation}
In the Fig. 2 we present the age estimates for the same two cases
as in Fig. 1 in the $\Lambda$ + DDM theory. This is to be compared with the
best estimate of the current age of the universe
for $\Omega_\Lambda$ given by
Eq.~(\ref{jedan}) is (Perlmutter et al.~1999)
\begin{equation}
\label{age}
t_0 = 14.9^{+1.4}_{-1.1} \times \frac{0.63}{h} \;{\rm Gyr}.
\end{equation}
In addition, it should be compared to the age of globular clusters
recently carefully measured with accounting for the revised
Hipparchos distance scale (Chaboyer et al. 1996, 1998). We
perceive that the estimates presented in Figure 2 are rather
significantly smaller from the result in Equation (\ref{age}),
although they have a correct correspondence limit $\Lambda =0$
of $t_0 \approx 12$ Gyr as in Sciama (1997). So low ages are
inconvenient from the point of view of globular cluster ages, as
well as our understanding of the most distant galaxies observed.

\begin{figure*}

{\includegraphics[height=9cm,angle=270]{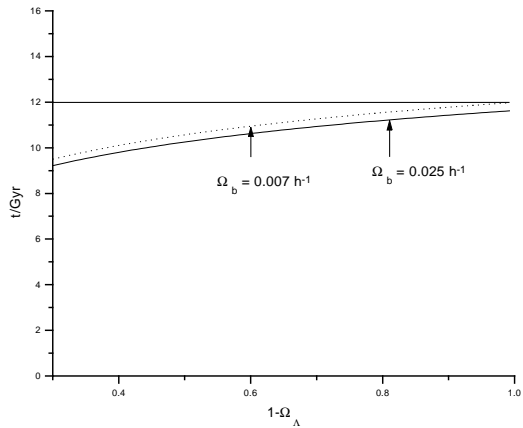}}


\caption{The ages predicted by $\Lambda$ + DDM flat cosmological model.
The same notation as in Figure 1 is used. \label{auto2}}
\end{figure*}

Considering the current trend in observational estimates of
cosmological parameters, the impact of cosmological constant on
parameter values in DDM theory is largely negative. Resulting
values of corrected parameters for $\Lambda \neq 0$ version of
Sciama's theory make the entire scheme less plausible. In that
respect, recent results of the EURD mission are highly indicative
of the observational verdict. This mission failed to observe the
emission of the dark sky at wavelengths slightly lower than 912
\AA, with the limit (95\% confidence) of only a third of the
predicted intensity.

The negative EURD result is not the only
indication of problems of the DDM theory. Recently, Maloney \& Bland-Hawthorn
(1999) calculated the flux from a full-neutrino halo and obtained:
$\phi\approx 2-3\times 10^5$ photons cm$^{-2}$ s$^{-1}$.
They find that the observed emission is much fainter:
$\phi\sim 10^4$ photons cm$^{-2}$ s$^{-1}$. The detrimental
consequences of this result for DDM cannot be remedied by
introduction of $\Lambda$, since the latter does not impact
galactic dynamics, and therefore the estimates of necessary amount
of dark matter in galactic haloes. Of course, one could always
assert that DDM is only a small ($\sim 10$\%) part of the dark
halo, most of it being in the form of baryons (i.e.\ MACHOs and/or
molecular clouds; see Fields, Freese and Graff 1998; Samurovi\'c, \'Cirkovi\'c
and Milo\v sevi\'c-Zdjelar 1999, also Sciama 2000). In that version, DDM is dominant only on
larger than galactic scales. However, the entire rationale of
the theory is undermined in this way, since there is no more any direct
connection between cosmology and the ISM physics, and the properties of
the decaying neutrino can not be constrained with remarkable precision
any more. The same criticism applies to the open DDM models (i.e.\ $\Omega_\nu \approx
\Omega < 1$) which requires even more fine-tuning, in particular
in view of the consequent properties of matter in galaxy clusters.

However, even the fine-tuned version of the theory fails if
confronted with negative results in particle experiments on
neutrino masses. Although recent results on the oscillations of
atmospheric neutrinos (Fu\-ku\-da et al.\ 1998) are sensitive only
to the mixing angles and mass difference $\Delta m^2$ between the
two neutrino flavors, the results are somewhat indicative in
suggesting rather low, probably sub-eV neutrino masses. Although
the DDM theory was correct in assuming neutrino masses---the first
empirical result in particle physics outside of the Standard
Model---only experiments currently in progress will show whether the
required neutrino masses are compatible with empirical limits.

Obviously, the simplicity and elegance of the original DDM theory
is lost with any complication such as discussed in the
present paper. Any attempt of bringing it in accordance with the
observational data must result, it seems, in more and more contrived
versions of the original beautiful idea. In this sense, we may compare
it with the classical steady state cosmological model of Bondi and Gold
(1948), as well as Hoyle (1948), which has been discredited in the
course of progress of observations, but which has had an epochal
impact on the very formation of modern cosmology (Kragh 1996).
In the same manner, Sciama's DDM theory, although it may be regarded
as disproved by now, has inspired and provoked an immense theoretical and
observational activity in astrophysics and cosmology. The results of these
efforts will certainly present its lasting legacy.

\section*{Acknowledgments}

The authors wholeheartedly thank  Vesna
Milo\v sevi\'c-Zdjelar for help in finding several important
references. S.S. acknowledges the financial support of the Abdus
Salam International Centre for Theoretical Physics, Trieste. This
research has made use of NASA's Astrophysics Data System Abstract
Service.

\end{document}